\documentstyle[preprint,pre,aps,epsf,times]{revtex}

\newcommand{\qq}{\mbox{\bf q}}
\newcommand{\Vol}{{\mbox{Vol}}}

\newcommand{\vv}{\mbox{\boldmath$v$}}
\newcommand{\xx}{\mbox{\boldmath$x$}}

\newcommand{\hT}{T}
\newcommand{\hq}{q}
\newcommand{\hDelta}{\Delta}
\newcommand{\hx}{x}
\newcommand{\hsigma}{\sigma}
\newcommand{\BGK}{\mbox{\tiny BGK}}
\newcommand{\MM}{\mbox{\tiny MM}}
\newcommand{\SW}{\tiny \mbox{SW}}

\newcommand{\otau}{\stackrel{\circ}{\tau}}

\newcommand{\dd}{\mbox{d}}
\newcommand{\mb}{\mbox{\tiny b}}

\draft
\begin{document}
\title{Boundary layer variational principles: A case study}
\author{Miroslav Grmela\\
\'Ecole Polytechnique de Montreal,  H3C 3A7 Montr\'eal (Qu\'ebec), Canada \\
Iliya V.\ Karlin
\thanks{Corresponding author. E-mail: ikarlin@ifp.mat.ethz.ch}\\
Swiss Federal Institute of Technology, Department of Materials, Institute of Polymers\\
ETH-Zentrum, Sonneggstr.\ 3, ML J 19, CH-8092 Z\"{u}rich,
Switzerland\\
Vladimir B.\ Zmievski\\
\'Ecole Polytechnique de Montreal,  H3C 3A7 Montr\'eal (Qu\'ebec), Canada } \maketitle
\begin{abstract}
Considering the model heat conduction problem in the setting of Grad's moment equations, we demonstrate a crossover in
the structure of minima of the entropy production within the boundary layer. Based on this observation, we formulate
and compare variation principles for solving the problem of boundary conditions in nonequilibrium thermodynamics.
\end{abstract}

\section{Introduction}

The goal of this paper is to study possibilities of formulating variational principles for
boundary conditions appearing in the extended thermodynamic systems, where the usual
locally conserved fields (the mass density, the momentum density, and the energy density)
are supplemented by various non-conserved fields such as extra stresses in rheology
\cite{Bird}, higher-order moments of the one-particle distribution function in moment
systems derived in kinetic theory of gases and plasmas, and many others.  In order to be
specific, we shall restrict our attention to the case of so-called extended thermodynamic
system, underpinned by Grad's moment method of the Boltzmann equation of rarefied gas
\cite{Grad}.  Since the seminal work of Grad \cite{Grad}, it is well known that the
stationary problems in moment equations is ill-posed.  Indeed, on physical grounds, it is
often unclear how to infer the values of the higher moments on the boundaries without a
more microscopic considerations.

In a situation where imposing boundary conditions is problematic or ill-posed, two main
directions in the search for formulations of the boundary conditions can be distinguished.
The first direction can be broadly characterized as a variational approach.  A typical and
quite well known representative of this strategy are so-called natural variational
formulations of stationary equations \cite{Veinberg}. This approach is widely used, in
particular, in numerical methods based on local minimization schemes, such as the finite
elements method \cite{Zienkewitz}.  Without going into any detail here, we mention that if
the solution can be written as a minimizer of a functional, then it is sometimes possible
to extend the solution from the bulk to the boundary, or to modify the functional in such
a way as to make this extension possible.  By doing so, the natural variational
formulations results in so-called natural boundary conditions.  Many
examples are given in the standard references on the finite elements method
\cite{Zienkewitz}.  It should be
also noticed that the physical significance of the boundary condition thus arising is
rarely addressed, especially in the case of extra fields without direct physical
interpretation.  The physics that is behind the behavior in the bulk may not be identical
with the physics that is behind the boundary conditions.  For example, new type of forces
arise often on boundaries.  It is thus possible that a direct extension to boundaries of
the potentials that are found to express the physics in the bulk is unrealistic.

The second strategy is based on an attempt to express the physics that takes place on the
boundaries.  Let us assume that as the result of the physical analysis one formulates a
coupled system of equations governing the time evolution in both the bulk and on
boundaries.  States on the boundary, or in a boundary layer, are described by values on
the boundary (or in boundary layers) of the fields chosen to describe states in the bulk
and possibly by some other fields defined only on boundaries or boundary layers.  Some of
the boundary state variables are fixed by an outside influence.  The rest of them, the
uncontrollable boundary state variables, evolve in time together with the bulk state
variables.  Let us assume that analysis of the time evolution equations shows that the
uncontrolled boundary state variables evolve faster than the bulk state variables and that
they approach, as the time goes to infinity, stationary values.  These asymptotically
reached stationary values of the boundary state variables are then the boundary conditions
that we look for.  We obtained them thus by solving the time evolution equations.  If in
addition, we are able to recognize in the analysis of the fast time evolution a Lyapunov
functional, then also this second strategy becomes a variational method.  This is because
the boundary conditions we look for are in such a case extremal values of the Lyapunov
functional.  It is important to emphasize that the way the variational functional is
introduced in this second strategy does not use the potential arising in the bulk, it does
not even use the assumption that such potential exists.  In fact, it is well known
\cite{Gross} that the time evolution in the bulk of driven systems can not be often
associated with any potential.  The potential introduced in the second strategy arises
from the time evolution of the boundary state variables and not from the time evolution in
the bulk.

The second strategy has been mentioned in Ref.  \cite{Grmela} as an illustration of a
general approach to the thermodynamics of driven systems.  The potential-driven time
evolution of boundary state variables have also been used in \cite{Shore} in the context
of the investigation of consequence of the stick-slip boundary conditions in flows of
polymeric liquids.  The authors of Ref.  \cite{Shore} do not discuss the physical
derivation of the boundary time evolution.  Also the potential is introduced in
\cite{Shore} completely phenomenologically.

Our study in this paper remains also on a phenomenological level.  We do not discuss the explicitly the boundary time
evolution, we are not therefore in position to recognize the pertinent potential in its analysis.  We have to use different
considerations in order to identify it.  Below, we shall follow a recent work of Struchtrup and Weiss \cite{SW} (see
also Ref.\ \cite{Mueller}).  Struchtrup and Weiss \cite{SW} proceed in three steps.

First, they suggest to consider the local entropy production $\sigma$ as a candidate for the potential that will
eventually determine the missing boundary conditions. While it is quite well known \cite{Gross} that $\sigma$ cannot
always be directly related to the time evolution in the bulk, it can still be relevant to the boundary conditions
(especially in the light of our expectation - based on the physical analysis sketched in the previous paragraph - that
the boundary time evolution can always be associated with a potential).

Second, having chosen $\sigma$, one has to ask the question how does this
potential depend on the boundary conditions. Struchtrup and Weiss \cite{SW}
answer this question as follows: First, they limit the analysis to stationary
solutions. Let the stationary solution
corresponding to a given boundary condition is found. The entropy production
$\sigma$, evaluated on the stationary solution, becomes a function of both the
bulk and the boundary state variables.

So far, we have arrived at a potential depending on the bulk and the boundary state variables. What remains is to make
the third step, and eliminate the dependence on the bulk variables. It is this third step where our analysis differs
from Refs.\ \cite{SW,Mueller}. It has been noticed in Ref.\ \cite{SW} that elimination of the bulk variables by
averaging the entropy production over the entire volume - and which eventually leads to the total entropy production
principle in a spirit of Glansdorff and Prigogine \cite{Prigogine,GP} - gives apparently wrong results in applications
to the boundary condition problem. Instead, a different, much more local analysis has been adopted in
\cite{SW,Mueller}. However, the physical significance of such modifications, as well as the physical reasons why the
global averaging out the bulk variables is not be working have not been addressed.

In this paper we address the question how physically meaningful
variational principles for boundary conditions can be constructed on the basis of
the entropy production by exploring more possibilities than those explored
in \cite{SW,Mueller}.
The intuitive idea behind our consideration is
that the additional variables in stationary problems often have a significance of a
description of the boundary layer (this description is greatly reduced, as compared to a
full kinetic equation).  By adopting this viewpoint, we study the question as to what
happens if the entropy production is considered not in the total volume of
the system but rather is localized to sufficiently thin boundary layers.  A physical
interpretation of our results is as follows:
If the domain of integration of the entropy
production is restricted to sufficiently thin boundary layers, the result of the
 type minimization suggests the optimal choice of the boundary
condition.  Moreover, shrinking the domain where the entropy production is sampled from
the whole bulk to the boundary layer reveals a behavior typical for a critical phenomena,
with the optimal value of the boundary condition appearing as a result of passing a
critical size of the layer.
Various features of this transition are studied, and plausible
realizations of the minimum principle are suggested.

\section{Entropy production in the boundary layer}

In the context of Grad's method \cite{Grad} and its variations, the state of the system is described by the locally
conserved fields $M(\xx,t)$ (the local density, momentum and energy), and a finite number of nonconserved fields,
$N(\xx,t)$ (nonequilibrium stress tensor, heat flux, fluxes thereof etc).  The fields $N$ are usually higher order
moments of the distribution function which gives a full description of the system at a more microscopic level of the
kinetic equation.  Grad's method reduces in a systematic way the description from the level of the kinetic equation for
the one-body distribution function to the level of a closed set of the moment equations involving only the fields $M$
and $N$.  The nonlinear coupled sets of equations in partial derivatives are generically referred to as Grad's moment
equations, and are given in many sources.  The original Grad's method \cite{Grad}, technically based on a Hermite
polynomial expansion of the one-body distribution function satisfying the Boltzmann kinetic equation, has been extended
and modified by many authors for various kinetic equations \cite{Kogan65,Mueller93,Levermore96}.  In particular, a
generalization of Grad's method to non-moment variables has been addressed in \cite{Karlin86,Karlin89,GK96}.  Examples
of Grad's moment equations will be considered in the next section.  Here we remind that, each Grad's moment system is
equipped with the function of the fields, $\sigma$, the local entropy production.  Function $\sigma$ is nonnegative and
equals to zero only at the local equilibrium, and it can be computed once the dissipative terms in the underlying
kinetic equation are specified (for example, once the Boltzmann collision integral is specified). The form of the
entropy production also depends on the version of Grad's method used in the derivation of moment equations.  In many
applications, the typical outcome for the entropy production is a quadratic form in the fields $N$ (this is valid for
small deviations from local equilibrium),

\begin{equation}
\label{ep} \sigma=\sum_{ij} (N_{i}-N_{i}^{\rm eq}(M))A_{ij}(M)(N_{j}-N_{j}^{\rm eq}(M)),
\end{equation}
where $N_i^{\rm eq}(M)$ are values of the nonconserved fields in the local equilibrium (in terms of the kinetic theory,
the latter is given by the local equilibrium distribution function which depends perimetrically only on the locally
conserved fields $M$, the standard example is the local Maxwell distribution function), and where $A_{ij}$ is the
positive semidefinite matrix, with matrix elements dependent on the functions $M$ and also on the details of particle's
interaction in the kinetic picture (scattering cross-sections, for example).

In order to solve the stationary version of Grad's moment equations for the time-independent fields, $M(\xx)$ and
$N(\xx)$, in a domain $U \in R^n$, with the boundary $\partial U$, a set of boundary conditions should be provided.  In
the typical situation, which we here assume, the boundary conditions for the locally conserved fields $M(\xx)$ are
known, and the question concerns only the additional fields $N(\xx)$.  To this end, we adopt the first two steps as
suggested by Struchtrup and Weiss \cite{SW}:  First, we consider the set of all possible solutions to stationary Grad's
equations with the fixed boundary conditions for the conserved fields, $M_{\mb}=M(\xx)|_{\partial U}$, and with various
boundary conditions for the nonconserved fields, $N_{\mb}=N(\xx)|_{\partial U}$. (In principle, other types of boundary
conditions could be addressed, including derivatives of either $M$ or $N$, but we shall not consider this option here).
Second, evaluating the local entropy production functional (\ref{ep}) on the configurations of the fields thus
obtained, we get a set of functions, $\sigma(\xx, N_{\mb})$, parameterized by the boundary condition data $N_{\mb}$.
Finally, in the third step, one has to eliminate the dependence on $\xx$, and to end up with a potential
$\Psi(N_{\mb})$ depending only on the boundary data $N_{\mb}$, and whose minima should suggest the choice of the
boundary condition.  It is this third step where we offer a more detailed analysis, as compared to Refs.\
\cite{SW,Mueller}.

Specifically, we introduce an additional structure into the domain $U$. Being inspired by the concept of the boundary
layer, we introduce a one-parametric family of subdomains $B_L$, where $L\ge 0$. Each $B_L$ (the test boundary layer)
is thought as a subdomain of the characteristic thickness $L$, attached to the boundary $\partial U$. For the test
boundary layer $B_L$, we consider the layer-averaged entropy production,
\begin{equation}
\label{Sigma_BL}
\Sigma_{B_L}(N_{\mb})=\frac{1}{\Vol (B_L)} \int_{B_L}\sigma(\xx,N_{\mb}) d\xx,
\end{equation}
where $\Vol (B_L)$ is the volume of the subdomain $B_L$.

The study of minimizers of the set of functionals (\ref{Sigma_BL}) for various
characteristic thicknesses $L$ used to define the boundary layer is the central point of
our paper.
A priori, it is clear that, if the thickness of the layer is taken large
enough, then we eventually come to the total bulk-averaged entropy production,

\begin{equation}
\label{GP_total}
\Sigma_{U}(N_{\mb})=\frac{1}{\Vol(U)}\int_{U}\sigma(\xx,
N_{\mb})d\xx.
\end{equation}

As it has been already demonstrated with explicit examples in Ref.\ \cite{SW},
minimization of the functionals (\ref{GP_total}) over the boundary data $N_{\mb}$ selects
the field configurations beyond a reasonable physical interpretation.  On the other hand,
if we go into the opposite direction, taking thinner test boundary layers, and if the
hypothesis about the fields $N$ as playing the most important role in the description of
the physical boundary layer is right, we might expect a crossover in the structure of the
minimizers of the functional (\ref{Sigma_BL}).  Specifically, we expect that at some value
$L_{\rm c}$, a local minimum will start appearing, and which would correspond to the
physically plausible value of the boundary condition $N_{\mb}$.  We further expect that
variations of these minimal values is not large for the entire interval
$L\in [0,L_{\rm c}]$.

This expectation is also motivated in part by the suggestion of Struchtrup and Weiss
\cite{SW} who postulated a much more local functional as compared to the total
bulk-averaged entropy production (\ref{GP_total}), namely, that correct configurations
should minimize the maximum of the local entropy production, thus, considering the
functional,

\begin{equation}
\label{SW}
\Psi_{\SW}(N_{\mb})=\max_{\xx \in U}\sigma(\xx, N_{\mb}).
\end{equation}

Our suggestion to study functionals (\ref{Sigma_BL}) which sample the entropy production more locally in space as
compared to the total entropy production (\ref{GP_total}) does not coincide with the "ultralocal" functional
(\ref{SW}), and results are not expected to be identical even in the one-dimensional cases considered below. It should
be stressed that a correct mathematical definition of the system of the test boundary layers $B_L$ requires more
restrictions but we here do not consider this point rigorously here. Finally, the notion of the boundary layer is
pertinent to the underlying kinetic theory where it can be computed in a few model settings \cite{CercignaniBook}.
However, it is not straightforward to incorporate these results into our considerations.

Notice that the above construction does not eliminate completely
the $\xx$-dependence from the local entropy production $\sigma(\xx, N_{\mb})$,
rather, it replaces such a dependence by a more transparent one-parametric dependence
on the thickness $L$. In principle, any functional $\Sigma_{B_L}$ for
$L\in [0, L_{\rm c}]$ can be regarded as a potential.
However,  in practice,  a priori estimates for  the characteristic value of $L$ are sometimes available. These values
can be dependent on boundary conditions for controllable and uncontrollable fields as well (see next section).
On the other hand, the set of the subdomains suggests a realization for the potential $\Psi(N_{\mb})$ which compares
the averaged entropy production within the layer $B_L$ with the averaged entropy production within the rest of the
bulk, $U\setminus B_L$,
\begin{equation}
\label{rest}
\Sigma_{U \setminus  B_L}(N_{\mb})=\frac{1}{\Vol (U \setminus  B_L)}
\int_{U \setminus  B_L}\sigma(\xx, N_{\mb}) \dd \xx.
\end{equation}
Namely, smoothness of transition from the boundary layer into the bulk suggests the outcome for the boundary condition
$N_{\mb}$ which guarantees that the difference between the  averaged entropy production in the boundary layer and the
averaged entropy production in the bulk is minimal. This results in a minimization of the potential,

\begin{equation}
\label{Psi1} \Psi_1 (N_b)=|\Sigma_{B_L}(N_{\mb})-\Sigma_{U \setminus B_L}(N_{\mb})|.
\end{equation}

Other variational principles can be constructed on similar grounds. In particular, if one expects that a variation of
the entropy production in the boundary layer is considerably higher than in the bulk, then the functional $\Psi_1$
(\ref{Psi1}) can be replaced with a functional involving only local measures of activity
\begin{equation}
\label{Psi2} \Psi_{2}(N_b)=\max_{\xx \in B_L} \sigma(\xx,N_{\mb})-\min_{\xx \in B_L}\sigma(\xx, N_{\mb}).
\end{equation}
Here $\min_{\xx \in B_L}\sigma(\xx, N_{\mb})$ represents an approximation to the bulk activity. In the next section we
shall test all this in the context of a model of heat transfer.

\section{One dimensional heat conduction problem}

Following Refs.\ \cite{SW,Mueller}, we consider here a
one-dimensional problem of a stationary heat transfer for
Boltzmann's gas at rest placed between two walls with fixed
temperatures. The system is
described by Grad's 14-moment equations. The set of field variables
includes  hydrodynamic fields $M$ (the mass density $\rho(\xx)$,
the average velocity $\vv(\xx)$, and the temperature $T(\xx)$), as
well as the additional variables $N$, which are functions of
higher moments: The stress tensor $\tau(\xx)$, the heat flux
$\qq(\xx)$, and one more scalar field,  $\Delta(\xx)$, which
corresponds to the fourth-order moment of the one-particle
distribution function,
$$\Delta(\xx)=\int_{R^3} [f(\vv,\xx)-f^{\rm eq}(M(\xx),\vv)]v^4 d\vv,$$
where $f^{\rm eq}$ is the local Maxwellian.

We further assume that Grad's 14-moment distribution function, $f_{14}(M(\xx), N(\xx),\vv)$, depends only on one
spatial variable $x$, and that the velocity dependence is symmetric with respect to rotations in the $(v_y,v_z)$ plane.
In this case, the average velocity vector, the traceless part of the stress tensor $\otau$, and the $y$-, and the
$z$-components of the heat flux vector, are equal to zero.  The 14-moment Grad's system reduces to the system of four
equations for the mass density $\rho(x)$, for the pressure $p(x)$, for the heat flux $q(x)$, and for the fourth moment,
$\Delta(x)$, and it reads \cite{SW,Mueller,Mueller93}:

\begin{eqnarray}
\label{GradGeneral}
\partial_xq&=&0, \nonumber \\
\partial_x p&=&0, \nonumber \\
\partial_x(\Delta+15\frac{p^2}{\rho})&=&-6\frac{1}{\xi_1(\rho,T)}q, \nonumber \\
\partial_x \frac{q p}{\rho}&=&-\frac{1}{28}\frac{1}{\xi_2(\rho,T)}\Delta.
\end{eqnarray}

Here the pressure $p(\xx)$ is related to the temperature, and the density as $p=(k/m)\rho T$, and the positive
coefficients $\xi_1$ and $\xi_2$ are the relaxation times, which can be functions of the density $\rho$ and the
temperature $T$.  Explicit form of parameters $\xi_{1,2}$ is determined by the collision model used in the
corresponding Boltzmann equation.  Following \cite{SW,Mueller} we consider two models:  the Bhatnagar-Gross-Krook
equation (BGK model) which gives constant relaxation times $\xi_1=\xi_2=\tau$, and the gas of Maxwell molecules (MM
model) which leads to the choice $\xi_1=\xi_2=2/3(\alpha \rho)$, where $\alpha$ is a constant. The local entropy
production for Grad's system (\ref{GradGeneral}) reads:

\begin{eqnarray}
\label{sigma} \sigma = \frac{m}{k}\frac{1}{pT}\left\{ \frac{2}{5}\frac{1}{\xi_1} \frac{1}{T}q^2+
\frac{1}{120}\frac{1}{\xi_2}\frac{m}{k}\frac{\Delta^2}{T^2} \right\}
\end{eqnarray}

We assume that the walls are placed at $x=0$ and $x=a$.  Taking
into account the fact that the pressure and the heat fluxes are constant, the
equations (\ref{GradGeneral}) require one additional boundary condition (in addition to
boundary conditions for the temperature $T(0)=T_0$, $T(a)=T_1$)
in order to fix either the heat flux $q$ or the
variable $\Delta$ at one of the boundaries.

Let us first consider the BGK model.
It proves convenient to introduce reduced variables,
\begin{eqnarray}
\label{indim} &&T'=\frac{T}{T_0}, \qquad \Delta'=\frac{\Delta}{p (k/m)T_0}, \qquad
 q'=\frac{q}{p(\frac{k}{m}T_0)^{1/2}},   \nonumber \\
 &&  \hx'=\frac{x}{a}, \qquad  \sigma'= \frac{a T_0}{p(\frac{k}{m}T_0)^{1/2}} \sigma
\end{eqnarray}
Reduced variables (\ref{indim}) are used elsewhere below, and we omit
primes in order to save notation.

In terms of variables (\ref{indim}), Grad's equations (\ref{GradGeneral}) for
the BGK model may be written,
\begin{eqnarray}
\label{GradBGK}
\partial_{\hx}(\hDelta+15\hT)&=&-6\frac{1}{K_{\BGK}} \hq, \nonumber \\
\partial_{\hx} \hT &=&-\frac{1}{28}\frac{1}{\hq K_{\BGK}} \Delta,
\end{eqnarray}
where $$K_{\BGK}=\left(\frac{kT_0}{m} \right)^{1/2}\frac{\tau}{a},$$ is Knudsen number. The local entropy production
(\ref{sigma}) for the BGK model takes the form,
\begin{equation}
\label{sigmaBGK}
\hsigma = \frac{1}{K_{\BGK}}\left\{ \frac{2}{5}\frac{\hq^2}{\hT^2}+
\frac{1}{120}\frac{\Delta^2}{\hT^3} \right\}.
\end{equation}

Equations (\ref{GradBGK}) are easily solved analytically to give
\begin{eqnarray}
\label{GradBGKsolution}
&&\hT (x)     =  \hT_0 +  W(\exp(-\hx/s)-1)-\frac{2}{5}\frac{\hq \hx}{K_{\BGK}},    \nonumber \\
&&\hDelta (x) =  - 15 W \exp(-\hx/s)+\frac{56}{5}\hq^2, \\
&& \qquad W=\frac{\hT_1-\hT_0 + 2\hq/(5K_{\BGK})}{\exp(-1/s)-1}, \nonumber \\
&& \qquad  s=-\frac{28}{15} \hq K_{\BGK}\nonumber
\end{eqnarray}
We assume  $\hT_1>\hT_0$, then the meaningful values of the heat flux $\hq$ are negative.  Exponential decay near
the cold boundary $\hx=0$ indicates the boundary layer, and the absolute value of $s$ represents its effective
thickness (note that if $\hq<0$ then $s>0$). Notice that this thickness depends both on the Knudsen number, and on the
yet unknown boundary condition $\hat{q}$.

As it has been suggested in the previous section, we study minima of the
one-parametric family of the layer-averaged entropy productions,

\begin{equation}
\label{hatSigma_L}
\Sigma_{L}(\hq)=\frac{1}{2L}\left(\int_{0}^{L} \hsigma(x,\hq) dx+
\int_{1-L}^{1} \hsigma(x,\hq) dx\right).
\end{equation}

For small Knudsen numbers, and small difference of wall temperatures, results can be compared with the analytical
estimate for $\hq$ drawn from the conventional Fourier law. In that case, as it follows from the Chapman-Enskog
solution \cite{Chapman}, $\hq=-(5/2)K_{\BGK}\partial_{\hx}\hT$. This allows to analytically estimate the heat flux as
$\hq^* \approx -(5/2)K_{\BGK}(\hT_1-T_0)$. In the test discussed below the following set of parameters has been used:
$\hT(1)=1.1$, $\hT(0)=1$, and $K_{\BGK}= 0.05$, which results in the analytical estimate, $\hq^*=-0.0125$ for the heat
flux, and $|s| \approx 10^{-3}$ for the characteristic thickness of the boundary layer corresponding to this estimate.

\begin{figure}
\unitlength1cm
\begin{picture}(16,5)
\put(0,-1){
\epsfxsize=8cm \epsfbox{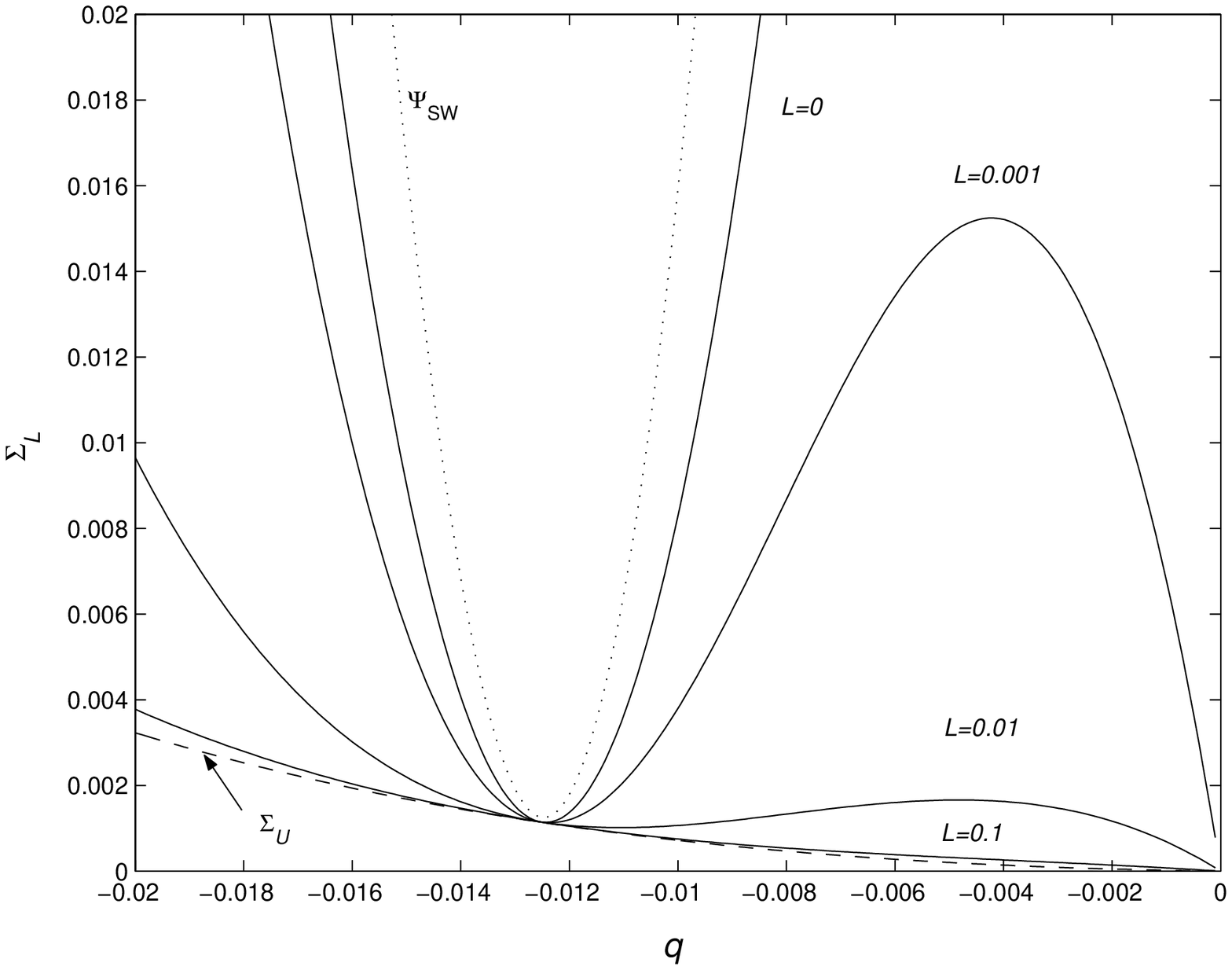}}
\end{picture}
\vspace*{1cm} 
\caption{Layer-averaged entropy production $\Sigma_{L}(q)$  (\protect\ref{hatSigma_L}) as a function of
the boundary condition $q$ for different layer widths $L$ in the BGK model with $K_{\BGK}=0.05$ and $T_1-T_0=0.1$.
Dashed line is the total bulk-averaged entropy production.}
\label{Fig1}
\end{figure}

First we compare functionals $\Sigma_{L} $ for various layer widths $L$.
Fig.\ \ref{Fig1} demonstrates the layer-averaged
entropy production $\Sigma_{L}(\hq)$ for different boundary
layer thickness $L$, including
the limit of the infinitely thin layer, $\lim_{L\to 0}\Sigma_{L}( \hq ) $,
as well as total entropy production, and
the functional of Struchtrup and Weiss (\ref{SW}).

We observe that, when $L$ varies from $1$ to $0$, there are two qualitatively different outcomes for the entropy
production $\Sigma_{L}$.  For $L$ larger than a crossover value $L_{\rm c}$, function $\Sigma_{L}(\hq)$
(\ref{hatSigma_L}) has one unphysical minimum $\hq=0$, which coincides with the minimum of the total bulk-averaged
entropy production (\ref{GP_total}).  The latter unphysical minimum has been already reported by Struchtrup and Weiss
\cite{SW}.  However, for $L \le L_{\rm c} $, function $\Sigma_{B_L}(\hq)$ (\ref{hatSigma_L}) demonstrates another local
minimum, $q_{\rm min}(L)$, although the unphysical minimum is still present.  As it is seen from Fig.  \ref{Fig1},
variations of the value $q_{\rm min}(L)$ is small within the interval $[0,L_{\rm c}]$, and all the values $q_{\rm
min}(L)$ are close to the analytical estimate $\hq^*$, on the one hand, and on the other hand, these values are close
to the minimizer of the function (\ref{SW}). This happens because the maximum of the local entropy production in this
and similar cases appears to be at the boundary, or within the boundary layer.  It is also remarkable that there is
invariant point where all curves $\Sigma_L(\hq)$ almost touch the curve corresponding to the total entropy
production.  This point is almost the same for any choice of $L$ and it is located very closely to the minimum of the
function $\Psi_1(q)$ (\ref{Psi1}).

\begin{figure}
\unitlength1cm
\begin{picture}(16,5)
\put(0,-1){
\epsfxsize=8cm \epsfbox{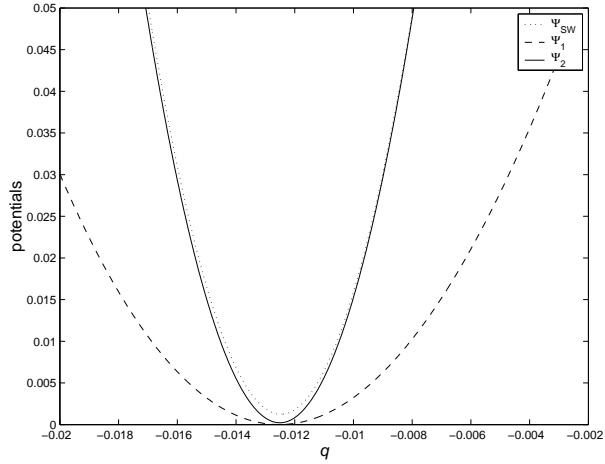}}
\end{picture}
\vspace*{1cm} 
\caption{Comparison of the potentials  $\Psi_1$ \protect(\ref{Psi1}) and $\Psi_2$ \protect(\ref{Psi2})
with the Struchtrup-Weiss potential $\Psi_{SW}$ \protect(\ref{SW}) in the BGK model with $K_{\BGK}=0.05$ and
$T_1-T_0=0.1$.} 
\label{Fig2}
\end{figure}

\begin{figure}
\unitlength1cm
\begin{picture}(16,5)
\put(0,-1){
\epsfxsize=8cm \epsfbox{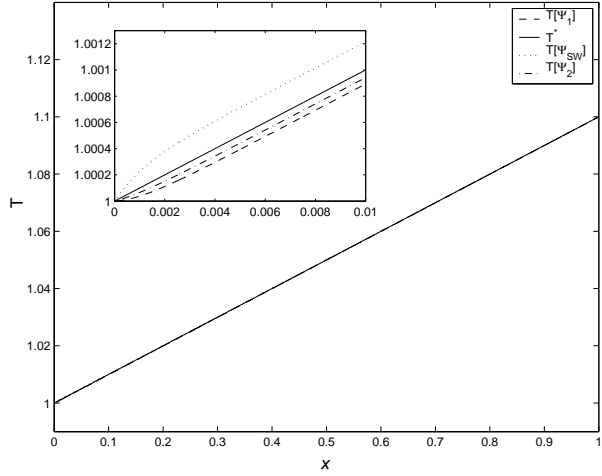}}
\end{picture}
\vspace*{1cm}
\caption{Profiles of the reduced temperature $\hT$ in the BGK model with $K_{\BGK}=0.05$ and $T_1-T_0=0.1$ corresponding to optimization with various
functionals.} 
\label{Fig3a}
\end{figure}

\begin{figure}
\unitlength1cm
\begin{picture}(16,5)
\put(0,-1){
\epsfxsize=8cm \epsfbox{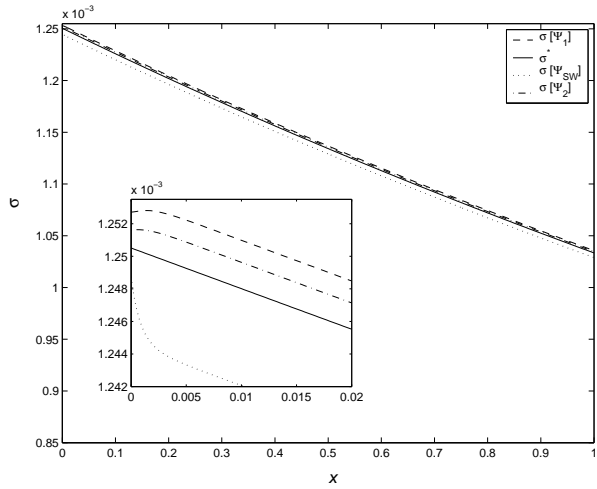}}
\end{picture}
\vspace*{1cm} 
\caption{Profiles of the reduced local entropy production
$\hsigma$ (b) in the BGK model with $K_{\BGK}=0.05$ and $T_1-T_0=0.1$ corresponding to optimization with various
functionals.}
\label{Fig3b}
\end{figure}

\begin{figure}
\unitlength1cm
\begin{picture}(16,5)
\put(0,-1){
\epsfxsize=8cm \epsfbox{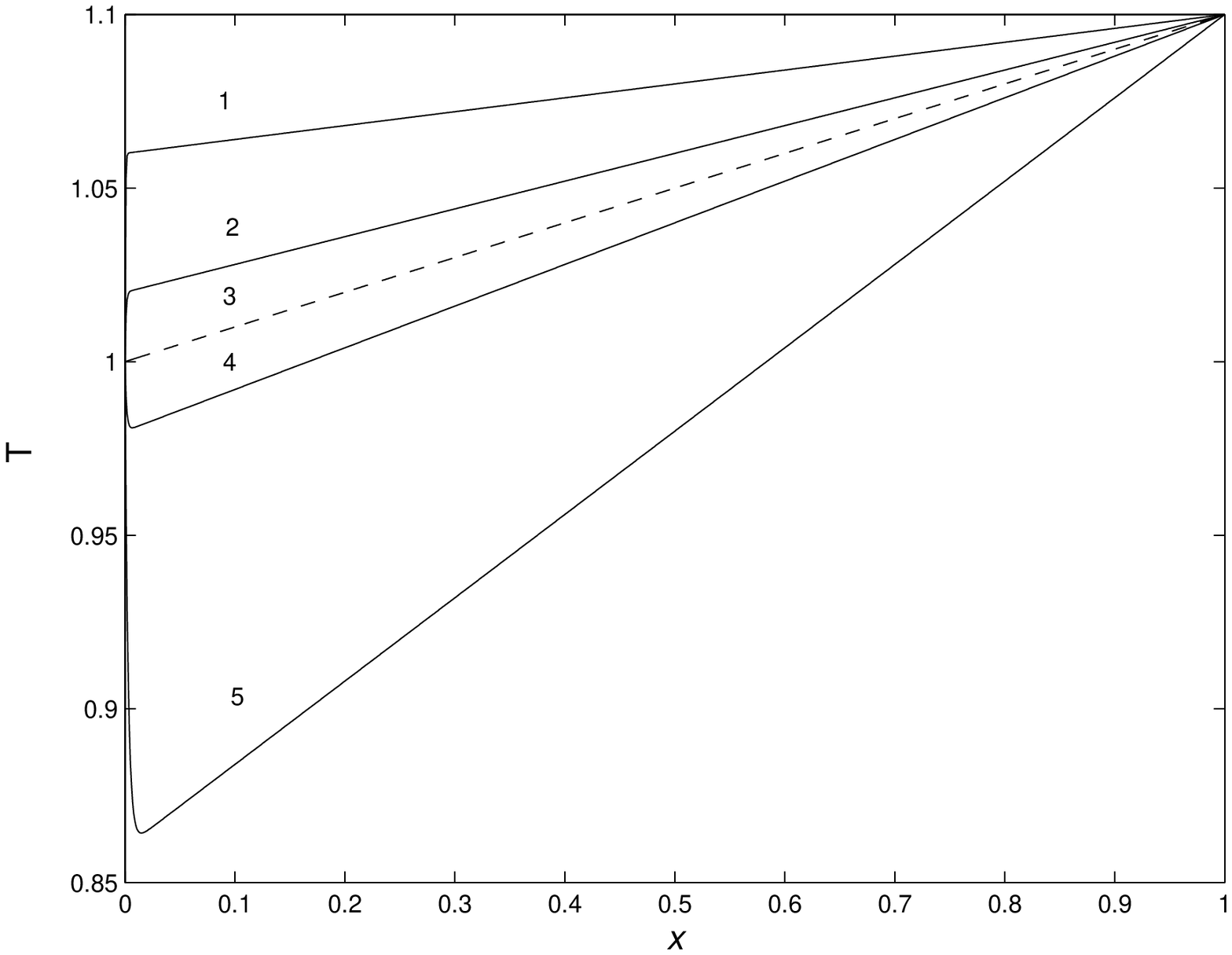}}
\end{picture}
\vspace*{1cm}
\caption{The reduced temperature  $\hT$ in BGK model with
$K_{\BGK}=0.05$ and $T_1-T_0=0.1$ for large deviations of boundary condition $\hq$ from its optimal value
$\hq^*=-0.0125$: curve 1 corresponds to   $\hq=-0.005$, curve 2 to $\hq=-0.01$,  curve 3 to $\hq=\hq^*$ (Fourier law),
curve 4  to  $\hq=-0.015$, curve 5 to $\hq=-0.03$.}
\label{Fig4a}
\end{figure}

\begin{figure}
\unitlength1cm
\begin{picture}(16,5)
\put(0,-1){
\epsfxsize=8cm \epsfbox{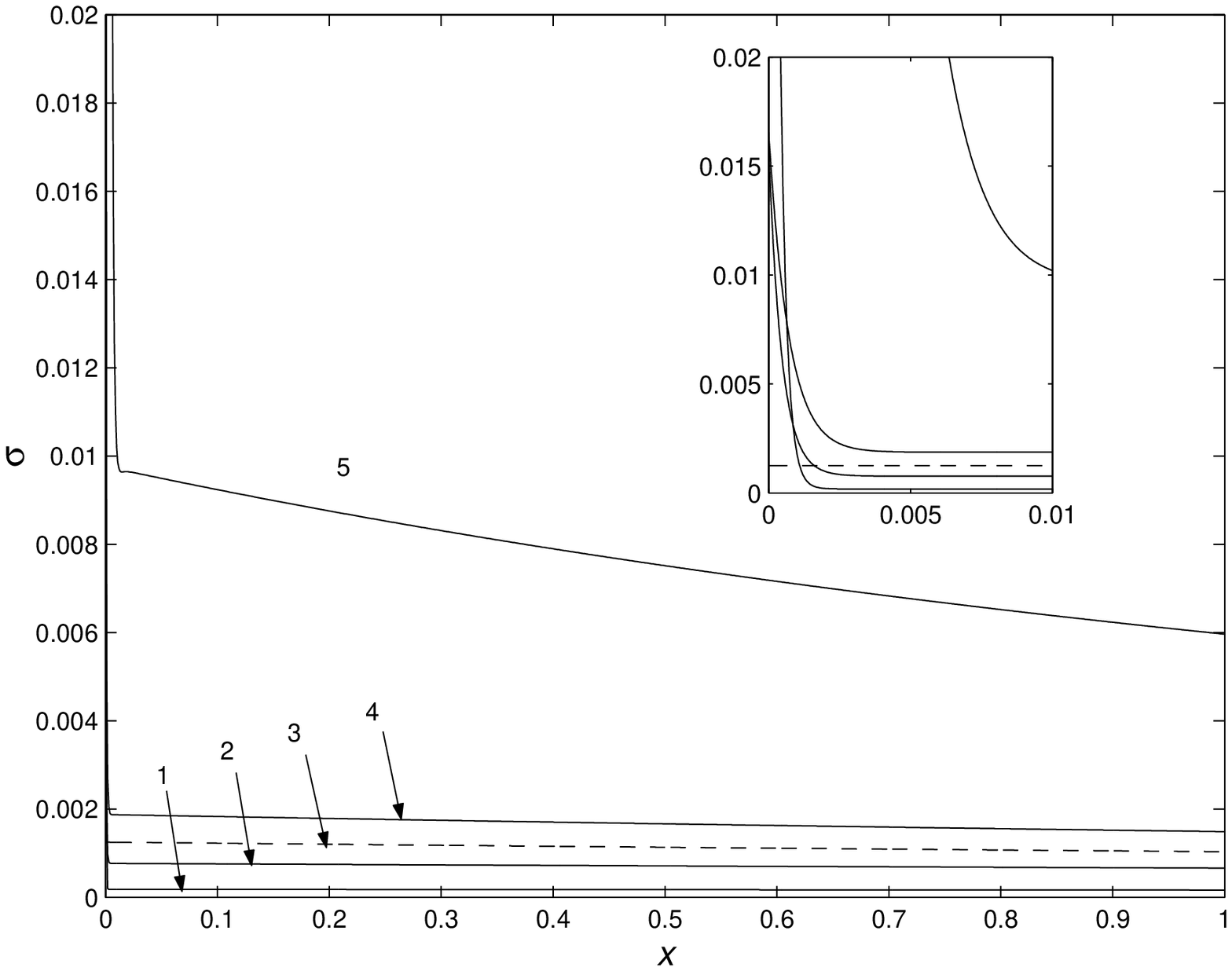}}
\end{picture}
\vspace*{1cm}
\caption{The reduced reduced local entropy production $\hsigma$ in BGK model with
$K_{\BGK}=0.05$ and $T_1-T_0=0.1$ for large deviations of boundary condition $\hq$ from its optimal value
$\hq^*=-0.0125$: curve 1 corresponds to   $\hq=-0.005$, curve 2 to $\hq=-0.01$,  curve 3 to $\hq=\hq^*$ (Fourier law),
curve 4  to  $\hq=-0.015$, curve 5 to $\hq=-0.03$.}
\label{Fig4b}
\end{figure}

Fig. \ref{Fig2} compares the three potentials, $\Psi_1$ (\ref{Psi1}), $\Psi_2$ (\ref{Psi2}), and $\Psi_{\tiny
\mbox{SW}}$ (\ref{SW}).
The value of boundary width $L$ in the definition of potentials $\Psi_1$ and $\Psi_2$ was fixed with help of estimate
$L(\hq)=s(\hq)$ (\ref{GradBGKsolution}) which is the function of boundary condition $\hq$.
The minima of these functionals correspond to the following values of $q$:
\begin{eqnarray}
q_{\min}[\Psi_1]&=&-0.012526\\\nonumber
q_{\min}[\Psi_2]&=&-0.012505\\\nonumber
q_{\min}[\Psi_{\SW}]&=&-0.012473
\label{comparison1}
\end{eqnarray}
All these values are very close to the analytical estimate $\hq^*=-0.012500$.
Notice that the estimate $\hq^*$ corresponds to the
most homogeneous profiles of the local entropy production, and also of
the
temperature (see Fig.\ \ref{Fig3a}, Fig.\ \ref{Fig3b}, and Fig.\ \ref{Fig4a}, \ref{Fig4b}).  Namely, one
observes that if the values $\hq$ are not in small vicinity of $\hq^*$ there is
an active domain near left wall $\hx=0$ where an exponential decay shows up.  It
is interesting to note that the boundary layer near right boundary does not have
any such activity, what is a consequence of the fact that at this boundary the
temperature flux is directed outward the bulk. In spite of slight deviations in
the results obtained with help of different potentials they give practically the
same temperature profiles.

\begin{figure}
\unitlength1cm
\begin{picture}(16,5)
\put(0,-1){ \epsfxsize=8cm \epsfbox{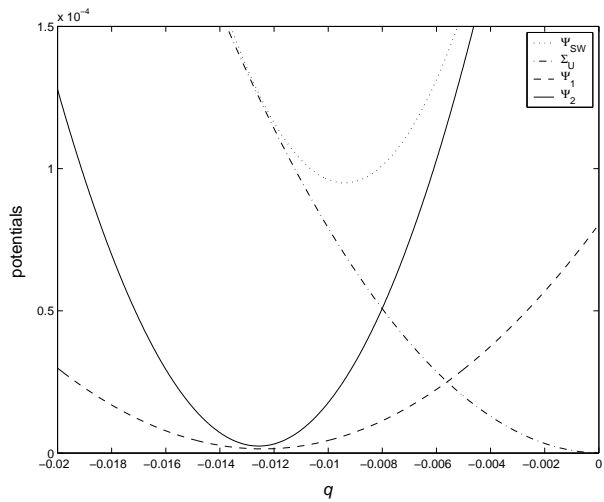}}
\end{picture}
\vspace*{1cm} 
\caption{ Functionals $\Psi_1$ \protect(\ref{Psi1}) and $\Psi_2$ (\protect\ref{Psi2}) as compared to
Struchtrup-Weiss potential $\Psi_{\SW}$ for the BGK model with moderate Knudsen number, $K_{BGK}=0.5$
($T_1-T_0=0.01$).} 
\label{Fig5}
\end{figure}

\begin{figure}
\unitlength1cm
\begin{picture}(16,5)
\put(0,-1){ \epsfxsize=8cm \epsfbox{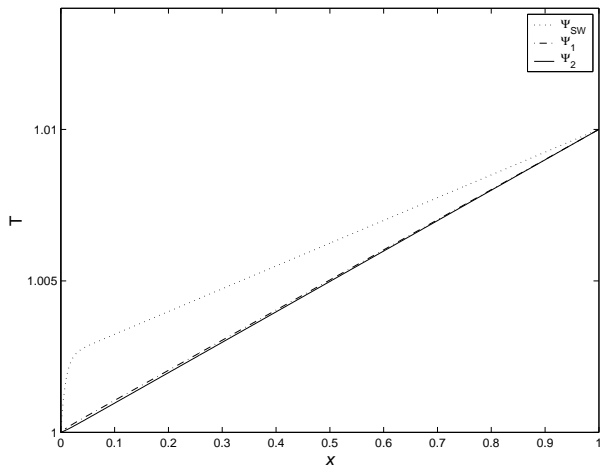}}
\end{picture}
\vspace*{1cm} 
\caption{Temperature profiles for moderate Knudsen number, $K_{\BGK}=0.5$ ($T_1-T_0=0.01$), in the BGK
model. } 
\label{Fig6}
\end{figure}

Although predictions based on all the three potentials, (\ref{Psi1}), (\ref{Psi2}), and (\ref{SW}), are close to each
other in the case of small Knudsen number, we have noticed considerable  divergency for larger Knudsen number. In order
to address this point, we have increased the value of the parameter $K_{\BGK}$, but have lowered the value for the
dimensionless temperature difference. Figures \ref{Fig5} and \ref{Fig6} correspond to the parameter set $K_{\BGK}=0.5$,
and $T_1-T_0=0.01$.
We then are able to qualitatively compare this result with the direct solutions to the linearized BGK equations
reported in the Ref.\ \cite{Bassanini}.  There is a clear indication that when the temperature difference between the
walls is sufficiently small the solution of BGK kinetic equations gives almost linear temperature profiles in the bulk
even for large Knudsen numbers.  Solution based on our variational principles confirms to this picture qualitatively.
However Figures 9 and 10 indicate that the Struchtrup-Weiss functional $\Psi_{\SW}$ point out the solution which is
considerably far from ``almost linear" unlike the case of small Knudsen numbers, which proves that our boundary layer
functional are more relevant to the problem of selection of boundary conditions.

Similar analysis has been performed for the model of Maxwell molecules.
In terms of variables (\ref{indim}), Grad's moment system for
the MM model reads:
\begin{eqnarray}
\label{GradMaxMol}
\partial_{\hx}(\hDelta+15\hT)&=&-4\frac{1}{K_{\MM}} \frac{\hq}{\hT}, \nonumber \\
\partial_{\hx} \hT &=&-\frac{1}{42}\frac{1}{K_{\MM}} \frac{\Delta}{\hq \hT},
\end{eqnarray}
where Knudsen number $K_{\MM}$ is,
$$
K_{\MM}=\frac{(kT_0/m)^{3/2}}{\alpha p a}.
$$
The local entropy production takes the form,

\begin{equation}
\label{sigma_mm}
\hsigma=\frac{1}{15 K_{\MM}}\left\{ 4\frac{\hq^2}{\hT^3}+
\frac{1}{12}\frac{\Delta^2}{\hT^4} \right\}.
\end{equation}

Because of the nonlinearity, equations (\ref{GradMaxMol}) were solved numerically. For small Knudsen numbers, and small
difference of the wall temperatures, the heat flux has been estimated as $\hq^* \approx-(15/4)K_{\MM}\hT'(\hT_1-T_0)$,
where $\hT'=(\hT_0+\hT_1)/2$. With this, the boundary layer is estimated as $L \approx \frac{42}{15}K_{\MM}\hT' |\hq|$.
Like for BGK model we have input the latter estimation into the expressions (\ref{Psi1}) and (\ref{Psi2}) in order to
completely specify the functions $\Psi_1(\hq)$ and $\Psi_2(\hq)$.
In the test presented below the following parameters were used: $K_{\MM}=0.05$, $T_0=1.0$, and $\hT_1=1.1$, which
results in the estimate $\hq^*\approx -0.01969$.

\begin{figure}
\unitlength1cm
\begin{picture}(16,5)
\put(0,-1){ \epsfxsize=8cm \epsfbox{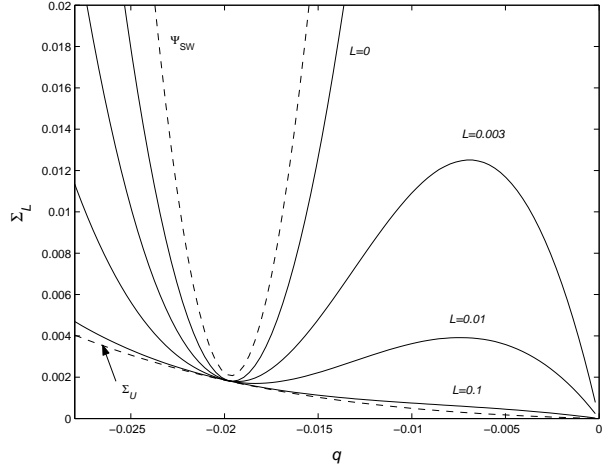}}
\end{picture}
\vspace*{1cm} 
\caption{Layer-averaged entropy production $\Sigma_{L}(q)$  (\protect\ref{hatSigma_L})  as a function of
the boundary condition $q$ for different layer widths $L$ in the Maxwell molecules model ($K_{\MM}=0.05$).
Dashed line is the total bulk-averaged entropy production.}
\label{Fig7}
\end{figure}

\begin{figure}
\unitlength1cm
\begin{picture}(16,5)
\put(0,-1){ \epsfxsize=8cm \epsfbox{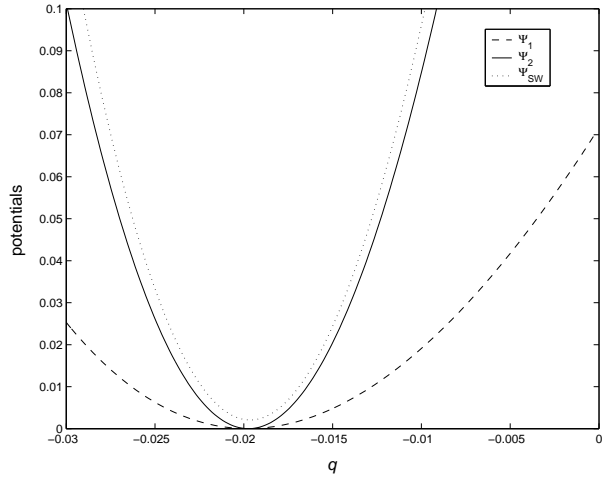}}
\end{picture}
\vspace*{1cm} 
\caption{Comparison of the potential  $\Psi_1$ \protect(\ref{Psi1}) and $\Psi_2$ (\protect\ref{Psi2})
with the Struchtrup-Weiss potential $\Psi_{SW}$ (\protect\ref{SW}) for the model of Maxwell molecules.} 
\label{Fig8}
\end{figure}

All results for the MM model are similar to those for the BGK model discussed above. Fig.\ \ref{Fig7} demonstrates the
crossover in the structure of the layer-averaged entropy production under variation of the layer width. Potentials
$\Psi_1$, $\Psi_2$ and $\Psi_{\SW}$ are compared in Fig.\ \ref{Fig8}. Corresponding minima of these potentials occur at
the following values of the heat flux:

\begin{eqnarray}
\hq_{\min}[\Psi_1]&=& -0.019777,\\\nonumber
\hq_{\min}[\Psi_2]&=&-0.019714,\\\nonumber
\hq_{\min}[\Psi_{\SW}]&=&-0.019643.
\label{comparison2}
\end{eqnarray}

\begin{figure}
\unitlength1cm
\begin{picture}(16,5)
\put(0,-1){ \epsfxsize=8cm \epsfbox{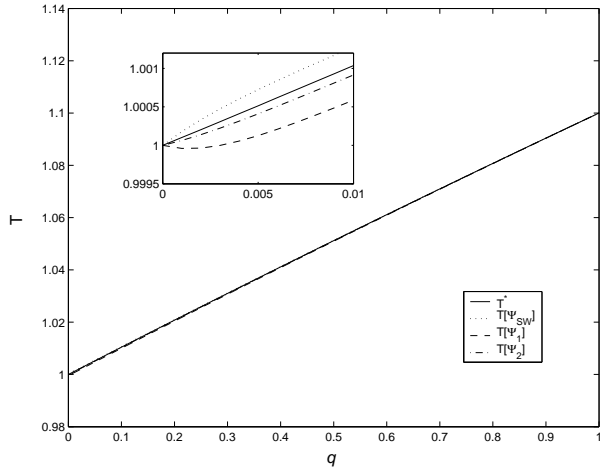}}
\end{picture}
\vspace*{1cm}
\caption{Profiles of the reduced temperature  $\hT$ in the model of
Maxwell molecules corresponding to minima of various potentials.} 
\label{Fig9a}
\end{figure}

\begin{figure}
\unitlength1cm
\begin{picture}(16,5)
\put(0,-1){ \epsfxsize=8cm \epsfbox{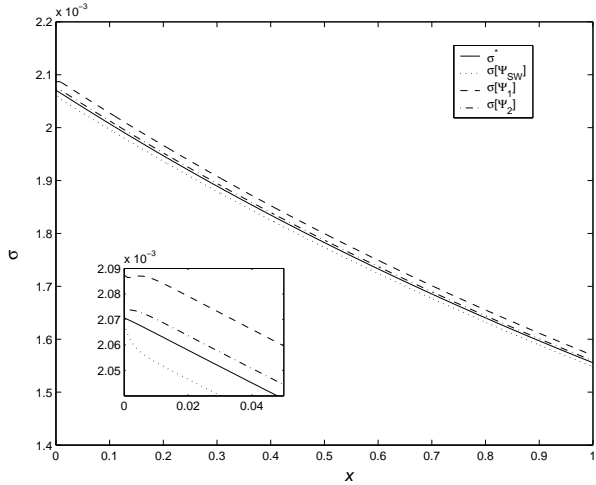}}
\end{picture}
\vspace*{1cm}
\caption{Profiles of the reduced  local entropy production $\hsigma$ in the model of
Maxwell molecules corresponding to minima of various potentials.} 
\label{Fig9b}
\end{figure}

All these values agree well with the estimate $\hq^*\approx -0.01969$. Notice that in both the BGK and the MM models,
potential $\Psi_2$ gives the result most close to the analytical prediction. Temperature and local entropy production
profiles are demonstrated in Fig.\ \ref{Fig9a} and Fig.\ \ref{Fig9b}.

\section{Conclusion and discussion}

In this paper, we have studied possibilities of introducing a variational principle for boundary conditions for Grad
moment equations.  Our approach is based on a systematic introduction of the boundary layer into a phenomenology of
variational principles. The approach has been tested for models of heat conduction suggested earlier. We have observed
that variation of the thickness of the domain taken to represent the boundary layer results in a crossover:  When
$L>L_{\rm c}$ then the minimum of the layer-averaged entropy production corresponds to the one predicted be the total
bulk-averaged entropy production. However, if $L<L_{\rm c}$, the second local minimum appears, and which corresponds to
the estimate close to the one resulting from the Struchtrup-Weiss minimax principle.  This crossover gives an
opportunity to define the boundary layer without restoring to more precise but also more elaborative microscopic
considerations. This observation has led us to variational principles which compare the average entropy production in
the boundary layer and in the bulk.
The results have been found in excellent agreement with analytical predictions.
The results of this study therefore make us confident in the usefulness of the
entropy production in the boundary layer for the problem of boundary conditions
in the extended thermodynamic systems.  The approach is computationally more
advantageous than the use of the minimax principle of \cite{SW} since it avoids
a computationally intensive operation of finding extrema of this entropy
production in entire volume, rather, it is based on a simple integral measure
and allow to use simplifications for small boundary layer width.

Finally, it should be stressed that, while the problem of boundary
conditions for moment equations (and, more broadly, for stationary
thermodynamic systems with additional fields) can be addressed
indeed through consideration of plausible minimum principles, the
complete understanding of those can be accomplished only in the
framework of dynamic approach to the boundary condition. This
point is left for future work.

\end{document}